\documentclass[
  preprint,
  superscriptaddress,
  showkeys,
  showpacs,
  preprintnumbers,
  nofootinbib,
  amsmath,amssymb,amsfonts,
  aps,
  prc,
  tightenlines,
  byrevtex
]{revtex4-1}

\usepackage{graphicx}
\usepackage{dcolumn}
\usepackage{bm}
\usepackage{hyperref}
\hypersetup{
  pdftoolbar=true,
  pdfmenubar=true,
  pdftitle={paper.pdf},
  pdfauthor={Christian H. Simon},
  colorlinks=true,
  linkcolor=blue,
  citecolor=blue,
  filecolor=blue,
  urlcolor=blue
}

\usepackage[all]{hypcap}

\begin{document}

\title{Interplay of anisotropies of momentum distribution and mean field in
heavy-ion collisions}

\author{Christian H. Simon}
\email{csimon@physi.uni-heidelberg.de}
\affiliation{Physikalisches Institut and Department of Physics and Astronomy,
  Ruprecht-Karls-Universit\"at Heidelberg, D-69120 Heidelberg, Germany}
\affiliation{National Superconducting Cyclotron Laboratory and Department of
  Physics and Astronomy, Michigan State University, East Lansing, Michigan
  48824, USA}
\author{Pawe\l~Danielewicz}
\email{danielewicz@nscl.msu.edu}
\affiliation{National Superconducting Cyclotron Laboratory and Department of
  Physics and Astronomy, Michigan State University, East Lansing, Michigan
  48824, USA}

\date{\today}

\begin{abstract}
\begin{description}
\item[Background] Two important parametrizations of momentum-dependent
nucleonic fields, proposed for the simulations of central heavy-ion collisions,
one by Gale \textit{et al.}\
[\href{http://dx.doi.org/10.1103/PhysRevC.35.1666}{Phys.\ Rev.\ C \textbf{35},
1666 (1987)}] and the other by Welke \textit{et al.}\
[\href{http://dx.doi.org/10.1103/PhysRevC.38.2101}{Phys.\ Rev.\ C \textbf{38},
2101 (1988)}], suffer
from practical limitations. The first gives rise to mean fields isotropic in
momentum, even when underlying momentum distributions are anisotropic, making
descriptions of early nonequilibrium stages of collisions unrealistic. The
second parametrization gives rise to anisotropic mean fields, but is
computationally expensive, because the mean field has to be computed separately
for every location of a nucleon in phase space, through folding.
\item[Purpose] Here we construct a parametrization of the nucleonic mean
field that yields an anisotropic mean field for an anisotropic momentum
distribution and is inexpensive computationally. To demonstrate the versatility
of our parametrization, we take the case of results from the parametrization by
Welke \textit{et al.}\
[\href{http://dx.doi.org/10.1103/PhysRevC.38.2101}{Phys.\ Rev.\ C \textbf{38},
2101 (1988)}] and attempt to approximate them.
\item[Method] In arriving at a suitable anisotropic mean-field potential, we
draw, on one hand, from the idea behind the parametrization of Gale \textit{et
al.}\
[\href{http://dx.doi.org/10.1103/PhysRevC.35.1666}{Phys.\ Rev.\ C \textbf{35},
1666 (1987)}], of a separable expansion of the potential energy, and, on the
other, from
the idea of a parallel expansion of the energy and mean field in anisotropy.
\item[Results] We show that using our novel parametrization we can
qualitatively and partially quantitatively reproduce the features of the
mean-field parametrization of Welke \textit{et al.}\
[\href{http://dx.doi.org/10.1103/PhysRevC.38.2101}{Phys.\ Rev.\ C \textbf{38},
2101 (1988)}].
\item[Conclusions] This opens up the possibility of exploring the effects of
mean-field anisotropy in collisions, without the penalty of computational cost
behind the folding parametrization.
\end{description}
\end{abstract}

\pacs{25.70.-z, 21.65.Mn, 21.60.-n}
\keywords{nuclear matter, heavy-ion collisions, transport theory,
          momentum anisotropy}

\maketitle
\section{\label{sec:intro}Introduction}
Heavy-ion collisions are complex events on account of many physical
effects competing in the dynamics. To effectively simulate the collisions,
simplifying assumptions and approximations need to be adopted. Fortunately,
excessive details are likely not essential for overall outcomes of those
simulations, given superposition of the effects over space and reaction history
and the fact that reaction observables tend to exhibit smooth variations with
characteristics of the reactions and measurement criteria.

Complexity and effective averaging over space and history in the collisions
make it tempting to simplify calculations at the level of numerical decisions.
However, this type of simplification has led to the common outcome for the
collision simulations, where simulations with the same or similar physical
assumptions yield\footnote{Community meetings have been dedicated to the issue,
in particular at ECT* Trento in 2003~\cite{Kol04} and 2006.} different
predictions for observables~\cite{Kol04}. Indeed, it is difficult to judge the
quality of approximations if an exact limit cannot be approached. With this,
it can be beneficial to adopt simplifications at the level of the theory for
the collisions in such a fashion that, on one hand, any sought interesting
physical effects can be captured and, on the other, the exact solution may be
approached in a systematic manner. The limit of exact solution can then serve
to validate the approximations, in addition to the conservation
laws~\cite{Len89}. In this context, we address here the formulation of momentum
dependence within the transport theory for heavy-ion collisions.

To provide the background, the heavy-ion collisions are commonly described in
simulations in terms of phase-space Wigner distributions~$f$, for nucleons and
other particles, that follow the Boltzmann-Uehling-Uhlenbeck (BUU)~\cite{Ueh33}
equation with nucleon optical potential~$U$. An efficient way of solving the
equation involves representing~$f$ in terms of test particles~\cite{Ber84},
$N_{\text{t}}$ per physical nucleon, with exact solution to the equations
approached for~$N_{\text{t}} \rightarrow \infty$. The collisions offer, in
particular, a unique opportunity for studying~$U$ at supranormal
densities~\cite{Dan00}. The interplay of the momentum dependence of the
nucleon optical potential~$U$ and collision observables turned out to be of
utmost importance in the heavy-ion-collision theory. Not only does momentum
dependence play a significant role in the
generation of collective flow, according to the transport
calculations~\cite{Pan93}, but it is also crucial for particle
production~\cite{Liq06}. In particular, to properly constrain the
nuclear compression modulus~$K$, defined as
\begin{equation}
K=p_{\text{F}}^2\frac{d^{2}(E/A)}{dp_{\text{F}}^{2}}\;,\label{eq:compress}
\end{equation}
where~$p_{\text{F}}$ is the Fermi momentum and~$E/A$ the binding energy per
nucleon, one must take the momentum dependence of nucleon-nucleon interactions
into account. In 1976, Blaizot \textit{et al.}~\cite{Bla76} showed that~$K$
could be inferred by measuring the energy of the isoscalar monopole resonance
in medium and heavy nuclei and arrived at the value of~$K=210\pm30
\,\text{MeV}$. Flow data, however, for quite some time seemed to be describable
by both a momentum-dependent ``soft'' eos ($K\approx 210\,\text{MeV}$) or a
momentum-independent ``hard'' eos
($K\approx 380\,\text{MeV}$)~\cite{Aic87,Gal87}. Eventually, though, it became
possible to decide on the momentum dependence with heavy-ion observables. Pan
and Danielewicz~\cite{Pan93} demonstrated that the best agreement between
sideward flow data and the results from transport calculations could be
obtained when applying a momentum-dependent soft eos.

Several approaches to the momentum dependence of the optical potential have
been put forward in the literature (see Ref.~\cite{Zha94} for a review article).
The particularly well-known parametrizations, utilized directly or indirectly in
BUU calculations, are the early
parametri\-zation by Gogny~\cite{Gog75}, the ansatz by Gale, Bertsch, and Das
Gupta~\cite{Gal87} and the one by Welke \textit{et al.}~\cite{Wel88}. In
Sec.~\ref{sec:theory}, the latter two parametrizations are discussed in more
detail, and the results by Welke \textit{et al.}\ serve as a reference for our
calculations. Following Landau quasiparticle theory~\cite{Noz99}, inherent in
these approaches are functional expressions for the potential energy
density~$V$ and for its functional derivative with respect to the
single-particle phase-space density~$f(\mathbf{r},\mathbf{p})$,
\begin{equation}
U=\left.\frac{\delta V}{\delta f}\right|_{\mathbf{p}},\label{eq:funcdiv}
\end{equation}
namely the nucleonic mean field (or optical potential)~$U$.

The important aspect of the formulation of Gale \textit{et al.}~\cite{Gal87},
of the momentum-dependent potential, is that the effort in
integrating~\cite{Len89} the mean-field part of the BUU equation grows linearly
with the test-particle number~$N_{\text{t}}$, paralleling the case of
momentum-independent~$U$, making it easy to approach the limit of an exact
solution. By contrast, the effort in integrating
the equation with the potential from Welke \textit{et al.}~\cite{Wel88} grows
as~$N_{\text{t}}^2$. The calculational convenience of Ref.~\cite{Gal87} comes,
though, at a price in that the optical potential~$U$ ends up being isotropic in
momentum space. However, if the phase-space density~$f$ is anisotropic in
momentum space, one may expect~$U$ to be anisotropic as well. The deficiency
of the formulation by Gale \textit{et al.}~\cite{Gal87} is likely to get more
and more serious the higher the energy of the collision and especially at the
early stages of the collision when~$f$ is most anisotropic. Here we show that
it is possible to extend the formulation~\cite{Gal87} to arrive at~$U$
anisotropic in momentum without scaling up the calculational effort as
in Ref.~\cite{Wel88}. To
demonstrate the flexibility of the new formulation for~$U$, we take on the
formulation by Welke \textit{et al.}\ and attempt to reproduce its results
for~$f$ representing typical situations in heavy-ion collisions.

Our work is based on the Master's~thesis~\cite{Sim11}. In our strategy, for the
sake of intrinsic consistency, we start out by considering the potential
energy~$V$ of a system anisotropic in momentum and derive~$U$
from~$V$~\eqref{eq:funcdiv}. We parametrize~$V$ taking into account, on one
hand, general physical expectations and, on the other, the need to carry out
efficiently BUU transport calculations, such as within the code by
Danielewicz~\cite{Dan00}, while ensuring the capability to explore different
anisotropies of~$U$. The construction of~$V$ involves introducing
separable interactions in $p$ space where different terms can represent
different spherical harmonics through which the anisotropy explicitly
enters the mean field. We attempt to make our approach suitable for different
stages of a heavy-ion collision. In comparing our results to those of Welke
\textit{et al.}, for partially equilibrated~$f$, we describe the anisotropy
of~$f$ in terms of an axial anisotropy parameter~$\varepsilon$. In addressing
the first stages of collisions, we consider the situation of two separated
Fermi spheres in momentum space, for projectile and target. Several efforts in
the literature~\cite{Gai99,Fuc03,Gai04}, with the goal of coping with the
early stage momentum anisotropies, specifically consider the latter type of
superposition.

The work is organized as follows. In Sec.~\ref{sec:theory}, we present the
models by Gale, Bertsch, and Das Gupta~(GBD) and Welke \textit{et al.}~(WPKDG).
The WPKDG model, requiring the determination of a three-dimensional integral for
every relevant phase-space location at every time step of a BUU calculation,
is computationally very costly. By contrast, GBD requires the determination of
integrals only for every relevant spatial location. In that section we also
introduce our own ansatz for~$V$ and~$U$, based on a spherical harmonics
expansion~(SHE). That ansatz also requires the determination of integrals only
for every relevant spatial location. Our SHE results for ground-state,
excited-state, and collision scenarios are represented in
Sec.~\ref{sec:results}, and our approach is tested there against
the WPKDG parametrization. The benefits of SHE for BUU calculations, in terms
of increased computational efficiency and reduced Monte Carlo noise, which are
disadvantages of WPKDG, are outlined in Sec.~\ref{sec:discuss}.

\section{\label{sec:theory}Theoretical considerations}
\subsection{\label{subsec:tA}Implicit anisotropy in mean fields}
The special feature of GBD is that it is formulated in the local frame where
the average nucleonic momentum~$\langle\mathbf{p}\rangle$ vanishes.
Alternatively, the nucleonic momenta may be specified relative
to~$\langle\mathbf{p}\rangle$ in another frame. This may be contrasted with
WPKDG.
Both models are based on Skyrme-type interactions~\cite{Sky59} for the
density-dependent part of the energy density and the mean field. In the case of
GBD, the potential energy density can be written as~\cite{Gal87}
\begin{eqnarray}
\label{eq:vgbd}
V_{\text{GBD}}(\rho(\mathbf{r}))=&~&\frac{A}{2}
\frac{\rho^2(\mathbf{r})}{\rho_0}+\frac{B}{\sigma +1}
\frac{\rho^{\sigma +1}(\mathbf{r})}{\rho^{\sigma}_0} \nonumber\\
&+&C\,\frac{\rho(\mathbf{r})}{\rho_0}\int d^{3}p'\,
\frac{f(\mathbf{r},\mathbf{p}')}{1+\left[\frac{\mathbf{p}'
-\langle \mathbf{p}\rangle}{\Lambda}\right]^2}\;.
\end{eqnarray}
Accordingly, following Eq.~(\ref{eq:funcdiv}), one obtains
the nucleonic mean field,
\begin{eqnarray}
\label{eq:ugbd}
U_{\text{GBD}}(\rho(\mathbf{r}),\mathbf{p})=&~&A\left(
\frac{\rho(\mathbf{r})}{\rho_0}\right)+B\left(
\frac{\rho(\mathbf{r})}{\rho_0}\right)^{\sigma} \nonumber\\
&+&\frac{C}{\rho_0}\int d^{3}p'\,\frac{f(\mathbf{r},\mathbf{p}')}
{1+\left[\frac{\mathbf{p}'-\langle \mathbf{p}\rangle}{\Lambda}\right]^2}
\nonumber\\
&+&\frac{C}{\rho_0}\frac{\rho(\mathbf{r})}
{1+\left[\frac{\mathbf{p}-\langle\mathbf{p}\rangle}{\Lambda}\right]^2}\;.
\end{eqnarray}
Note that~$f$---where not otherwise stated---represents the single-particle
phase-space density, normalized according to~$\rho(\mathbf{r})=\int
d^{3}p\,f(\mathbf{r},\mathbf{p})$.

Within WPKDG, the energy density is parametrized as~\cite{Wel88}
\begin{eqnarray}
\label{eq:vwelke}
V_{\text{WPKDG}}(\rho(\mathbf{r}))=&~&\frac{A}{2}\frac{\rho^2(\mathbf{r})}
{\rho_0}+\frac{B}{\sigma +1}\frac{\rho^{\sigma +1}(\mathbf{r})}
{\rho^{\sigma}_0} \nonumber\\
&+&\frac{C}{\rho_0}\int\int d^{3}p\> d^{3}p'\,\frac{f(\mathbf{r},
\mathbf{p})f(\mathbf{r},\mathbf{p}')}{1+\left[\frac{\mathbf{p}-\mathbf{p}'}
{\Lambda}\right]^2}\;.~~
\end{eqnarray}
This leads to the mean field of the form
\begin{eqnarray}
\label{eq:uwelke}
U_{\text{WPKDG}}(\rho(\mathbf{r}),\mathbf{p})=&~&A\left(\frac{\rho(\mathbf{r})}
{\rho_0}\right)+B\left(\frac{\rho(\mathbf{r})}{\rho_0}\right)^{\sigma}
\nonumber\\
&+&2\frac{C}{\rho_0}\int d^{3}p'\,\frac{f(\mathbf{r},\mathbf{p}')}
{1+\left[\frac{\mathbf{p}-\mathbf{p}'}{\Lambda}\right]^2}\;.
\end{eqnarray}
Inherent in both models are the five parameters: $A$, $B$, $\sigma$, $C$,
and~$\Lambda$. From those parameters, $\sigma$ is particularly strongly tied to
the incompressibility~$K$, while~$\Lambda$ determines the high-$p$ behavior
of~$U$. To constrain the parameters, a demand is placed that the energy per
nucleon minimizes with~$E/A=-16\,\text{MeV}$ at the saturation
density~$\rho_0=0.16\,\text{fm}^{-3}$. For both parametrizations, GBD and
WPKDG, the incompressibility is chosen equal to~$K\simeq 215\,\text{MeV}$, and
the effective mass ratio at saturation momentum is~$m^{*}/m\simeq0.7$. For
WPKDG,
specific values of the potential at~$\rho_0$ are~$U(\rho_0,p=0)\simeq-75
\,\text{MeV}$ and~$U(\rho_0,p^{2}/2m\simeq 300\,\text{MeV})=0$.
The~GBD parametrization may be considered an approximation to WPKDG, and this
issue has been explored to a degree in Ref.~\cite{Wel88}. With SHE we want to
develop a qualitatively better approximation to WPKDG, while still keeping the
method computationally inexpensive, as~GBD.
%%%%%%%%%%%%%%%%%%%%%%%%%%%%%%%%%%%%%%%%%%%%%%%%%%%%%%%%%%%%%%%%%%%%%%%%%%%%%%%
\subsection{\label{subsec:tB}Explicit anisotropy in mean fields}
The essential idea of SHE, borrowed from GBD~\cite{Gal87}, is that of a
separable representation for the energy functional. A separable representation,
with a limited number of terms, could be adjusted to represent results from
microscopic theory, obtained for different nonequilibrium situations, such as
Dirac-Brueckner~\cite{Fuc93}. In lieu of a microscopic theory, we take the
intuitively appealing WPKDG parametrization and examine to what extent we can
reproduce its results with SHE. If we can be successful here, we may be
successful with a microscopic theory as well. Otherwise, following a
phenomenological strategy, as common in heavy-ion collisions, we may use SHE as
the starting point and see whether collision observables can constrain the
energy functional constructed within SHE.
Besides the separable part of the energy, generating momentum dependence
of~$U$, we employ a $\rho$-dependent part of the Skyrme form, such as in GBD or
WPKDG. When aiming at a minimal number of terms in the separable expansion,
while retaining the capability of describing mean fields anisotropic in
momentum, our energy functional in SHE consists of Skyrme and separable scalar
and tensorial quadrupole terms:
\begin{eqnarray}
\label{eq:vsep}
V_{\text{SHE}}(\rho(\mathbf{r}))=&~&\frac{A}{2}\frac{\rho^2(\mathbf{r})}
{\rho_0}+\frac{B}{\sigma +1}\frac{\rho^{\sigma +1}(\mathbf{r})}
{\rho^{\sigma}_0} \nonumber\\
&+&\frac{C}{\rho_0}\left[~\left(\int d^{3}p\,\frac{f(\mathbf{r},\mathbf{p})}
{1+\left[\frac{\mathbf{p}}{\Lambda_{\text{iso}}}\right]^2}\right)^2\right.
\nonumber\\
&~&~~~~~+\left.\sum_{\alpha,\beta}T^{\alpha\beta}(\mathbf{r})\,T^{\alpha\beta}
(\mathbf{r})\right].
\end{eqnarray}
We do not include a dipole separable term, which we expect to be small in the
center of mass. A dipole term is accounted for in the relativistic approach of
Ref.~\cite{Mar94}.
To demonstrate the robustness of such a representation we
attempt to show that the results from the folding model WPKDG can be well
approximated within our separable model, including the reproduction of
anisotropies of~$U$. For simplicity, we take both the scalar and the tensor
terms as symmetric in two single-particle quantities, presuming dominance of
two-body interactions. Inclusion of a tensor term makes our parametrization
different from GBD, scalar in our terminology. However, we take the form of the
single-particle factor to be the same as in GBD~\cite{Gal87}.
In the above, $T^{\alpha\beta}$ is a second-rank Cartesian
tensor, motivated by the expansion of anisotropies into
spherical harmonics~\cite{App89,Dan05}:
\begin{equation}
T^{\alpha\beta}(\mathbf{r})\equiv\int d^{3}p\,\,c(p)\,f(\mathbf{r},
\mathbf{p})\,(p^{\alpha}p^{\beta}-\frac{1}
{3}\,p^{2}\delta^{\alpha\beta})\;. \label{eq:tab}
\end{equation}
An important feature is its tracelessness,
\begin{equation}
T^{xx}+T^{yy}+T^{zz}=0\;. \label{eq:trace}
\end{equation}
For the functional form of~$c(p)$ in Eq.~(\ref{eq:tab}) we choose
\begin{equation}
c(p)=\frac{1}{p^2+\Lambda_{\text{aniso}}^2}\;. \label{eq:cp}
\end{equation}
Note the parameters~$\Lambda_{\text{iso}}$ in the scalar part
of Eq.~(\ref{eq:vsep}) and~$\Lambda_{\text{aniso}}$  in the tensorial part,
which we introduced in addition to the ones inherited from GBD and WPKDG.

According to Eq.~(\ref{eq:funcdiv}), the optical potential is found by taking
the functional derivative of Eq.~(\ref{eq:vsep}) with respect to~$f$. Owing to
the directionality involved and approximate axial symmetry of the phase-space
density in heavy-ion collisions, the off-diagonal elements of Eq.~(\ref{eq:tab})
are negligible in practice, reducing the summation over~$\alpha,\beta$ to
diagonal elements only. Here the tracelessness~(\ref{eq:trace}) of the tensor
comes into play, allowing us to express all elements in terms of~$T^{zz}$ when
dependence on transverse direction is weak. Upon taking a functional derivative
of Eq.~(\ref{eq:vsep}) and transforming to spherical coordinates, we get
\begin{eqnarray}
\label{eq:usep}
U_{\text{SHE}}(\rho(\mathbf{r}),\mathbf{p})=&~&A\left(\frac{\rho(\mathbf{r})}
{\rho_0}\right)+B\left(\frac{\rho(\mathbf{r})}{\rho_0}\right)^{\sigma}
\nonumber\\
&+&\frac{C}{\rho_0}\Bigg[~\frac{2}{1+\left[\frac{\mathbf{p}}
{\Lambda_{\text{iso}}}\right]^2}\int d^{3}p'\,\frac{f(\mathbf{r},\mathbf{p}')}
{1+\left[\frac{\mathbf{p}'}{\Lambda_{\text{iso}}}\right]^2} \nonumber\\
&~&~~+\sqrt{\frac{16\pi}{5}}\,c(p)\,p^2\,Y_{20}(\vartheta)\,T^{zz}
(\mathbf{r})\,\Bigg].
\end{eqnarray}
Here we used the definition of the spherical harmonic of degree~$2$
and order~$0$,
\begin{equation}
Y_{20}(\vartheta)=\sqrt{\frac{5}{16\pi}}\,\left(3\cos^2(\vartheta)-1\right).
\label{eq:y20}
\end{equation}
Therewith, the SHE optical potential~(\ref{eq:usep}) depends on anisotropy
explicitly. The decisive last term vanishes for isotropic~$f$ because~$T^{zz}$
is zero for such cases by construction.

\section{\label{sec:results}Results}
%%%%%%%%%%%%%%%%%%%%%%%%%%%%%%%%%%%%%%%%%%%%%%%%%%%%%%%%%%%%%%%%%%%%%%%%%%%%%%%
\subsection{\label{subsec:rA}Ground-state properties}
As in GBD and WPKDG, we demand that~$E/A$ for SHE minimizes at~$\rho_0=0.16\,
\text{fm}^{-3}$ with the value of~$E/A=-16\,\text{MeV}$. For consistency with
the other parametrizations, we further demand that~$K\simeq 215\,\text{MeV}$,
and~$m^{*}/m\simeq0.7$ at the Fermi momentum of~$p_{\text{F}}^{(0)}=263\,
\text{MeV/c}$ at~$\rho_0$.
Otherwise, it would be natural to require that SHE reproduced the momentum
dependence of WPKDG for the ground-state phase-space density,
\begin{equation}
f(\mathbf{r},\mathbf{p})=\frac{g}{(2\pi\hbar)^3}\,
\theta(p_{\text{F}}^{(0)}-\left|\mathbf{p}\right|)\;, \label{eq:1fs}
\end{equation}
where the degeneracy factor is~$g=4$.
In fact, by adjusting the parameters of SHE the momentum dependencies
at~$\rho_0$ can be made nearly indistinguishable. However, then differences
between SHE and WPKDG may be excessive in some nonequilibrium situations.
Correspondingly, we adopt a compromise choosing parameters so that the results
have a similar appearance under different circumstances but can exhibit
quantitative discrepancies.
\begingroup
%\squeezetable
\begin{table*}
\caption{\label{tab:param}Parameters in GBD, WPKDG and SHE models.}
\begin{ruledtabular}
\begin{tabular}{cccccccc}
Model & A [MeV] & B [MeV] & C [MeV] & $\sigma$ & $\Lambda$ &
$\Lambda_{\text{iso}}$ & $\Lambda_{\text{aniso}}$\\
GBD & $-144.9$ & $203.3$ & $-75.0$ & $7/6$ & $1.5\,p_{\text{F}}^{(0)}$ & --- &
---\\
WPKDG & $-110.44$ & $140.9$ & $-64.95$ & $1.24$ & $1.58\,p_{\text{F}}^{(0)}$ &
--- & ---\\
SHE & $-110.63$ & $132.17$ & $-53.04$ & $1.27$ & --- &
$1.98\,p_{\text{F}}^{(0)}$ & $1.41\,p_{\text{F}}^{(0)}$\\
\end{tabular}
\end{ruledtabular}
\end{table*}
\endgroup

\begin{figure}
\includegraphics{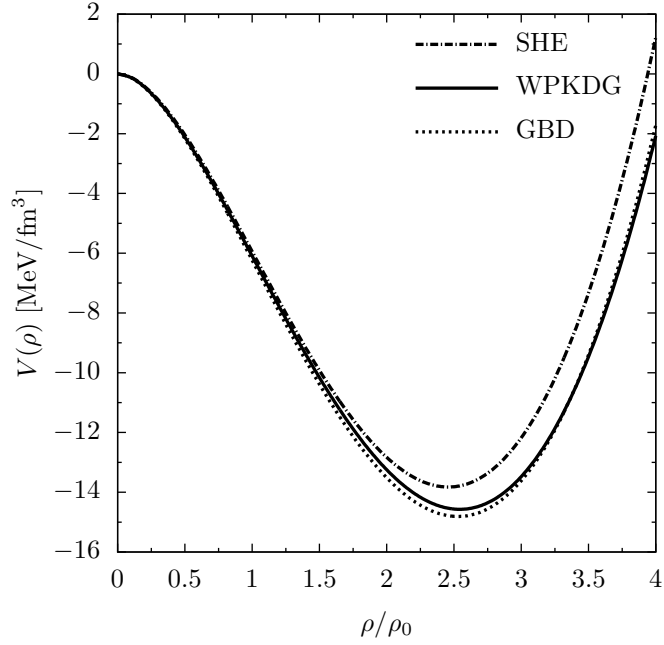}
\caption{\label{fig:cv1}Ground-state potential energy density as a function of
density of matter.}
\end{figure}

\begin{figure}
\includegraphics{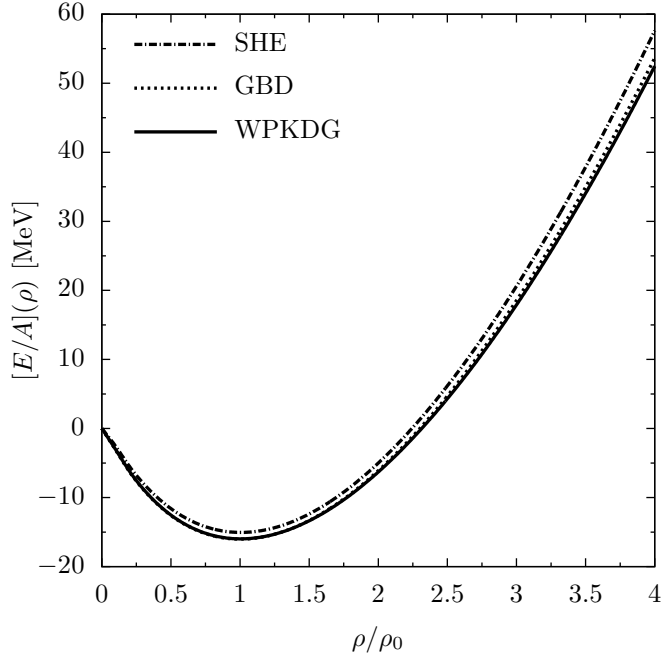}
\caption{\label{fig:ce1}Ground-state energy per nucleon as a function of density
of matter.}
\end{figure}

\begin{figure}
\includegraphics{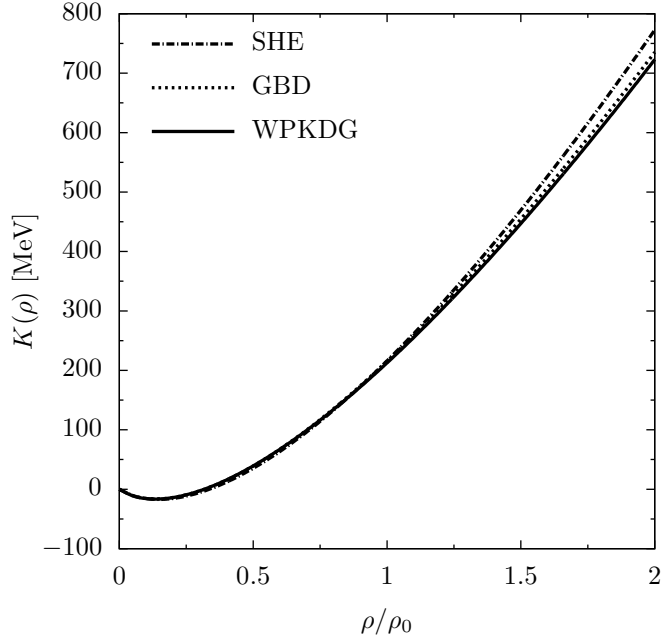}
\caption{\label{fig:ck1}Ground-state compressibility as a function of density of
matter.}
\end{figure}

\begin{figure}
\includegraphics{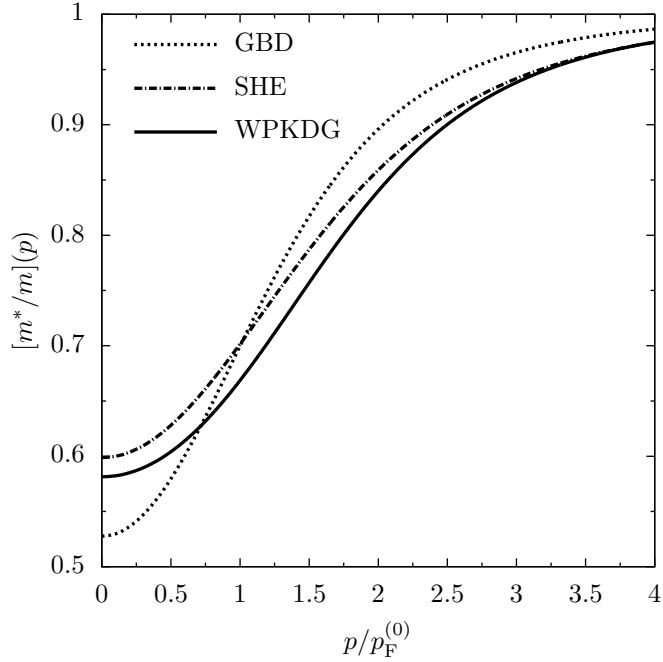}
\caption{\label{fig:cm1}Ground-state effective mass ratio as a function of
momentum, at normal density~$\rho_0$.}
\end{figure}

\begin{figure}
\includegraphics{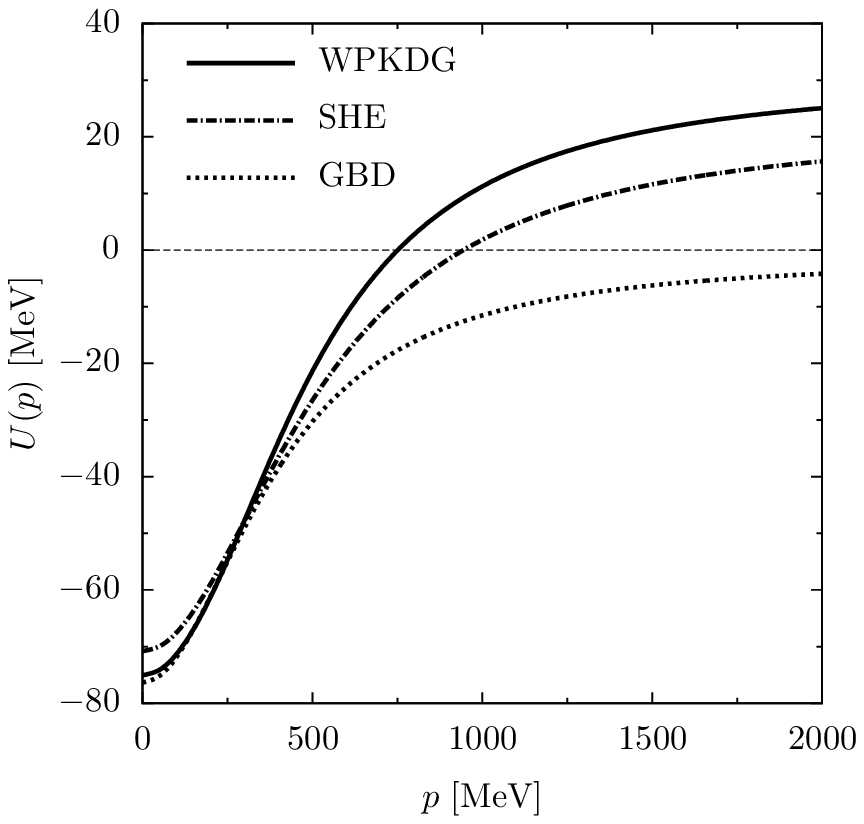}
\caption{\label{fig:cu1}Ground-state optical potential as a function of
momentum, at normal density~$\rho_0$.}
\end{figure}

At zero temperature, calculations of energy and optical potential can be
largely done analy\-tically. A couple of integrals useful for the purpose are
provided in the Appendix. With spatial density of kinetic energy given by
\begin{equation}
\left\langle\frac{p^2}{2m}\right\rangle=\int d^{3}p\,\frac{p^2}
{2m}f(\mathbf{r},\mathbf{p})\;, \label{eq:p2m}
\end{equation}
the energy per nucleon is
\begin{equation}
E/A=\int d^{3}r\left[\left\langle\frac{p^2}{2m}\right\rangle +V[f]\right]/A\;.
\label{eq:binding}
\end{equation}
From the energy, the incompressibility~$K$ is calculated and, from~$U$, the
effective mass ratio is determined,
\begin{equation}
\frac{m^{*}}{m}=\frac{p}{m}\left(\frac{d}{dp}\left[\frac{p^2}
{2m}+U(\rho(\mathbf{r}),\mathbf{p})\right]\right)^{-1}.\label{eq:m1}
\end{equation}
The parameters for SHE, as well as GBD and WPKDG, are provided in
Table~\ref{tab:param}. Figure~\ref{fig:cv1} shows the potential energy density
for the three interaction parametrizations and Fig.~\ref{fig:ce1} displays the
energy per nucleon. Next, Fig.~\ref{fig:ck1} shows the incompressibility as a
function of density~$\rho$ and Fig.~\ref{fig:cm1} shows the effective mass
ratio at~$\rho_0$, as a function of momentum. Finally, Fig.~\ref{fig:cu1} shows
the optical potential as a function of momentum at the normal density. As we
stated, by adjusting the parameters in SHE we could obtain an excellent
agreement between SHE and WPKDG in Figs.~\ref{fig:cm1} and~\ref{fig:cu1}, but
at the cost of poorer agreement in other situations.
%%%%%%%%%%%%%%%%%%%%%%%%%%%%%%%%%%%%%%%%%%%%%%%%%%%%%%%%%%%%%%%%%%%%%%%%%%%%%%%
\subsection{\label{subsec:rB}Excited nuclear matter}
\begin{figure}
\includegraphics{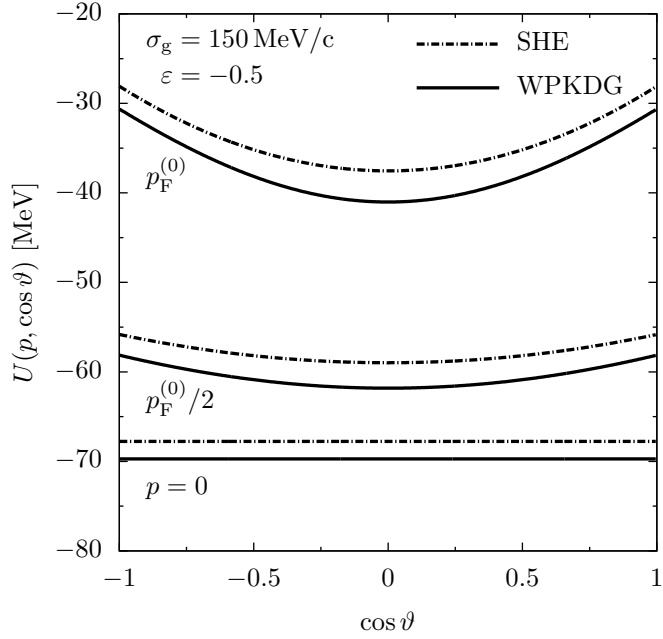}
\caption{\label{fig:cuga-05}Optical potential for an anisotropic momentum
distribution~(\ref{eq:gauss}) with~$\varepsilon=-0.5$ at~$\rho=2\rho_0$, in the
local c.m., for different indicated momenta~$p$.}
\end{figure}

\begin{figure}
\includegraphics{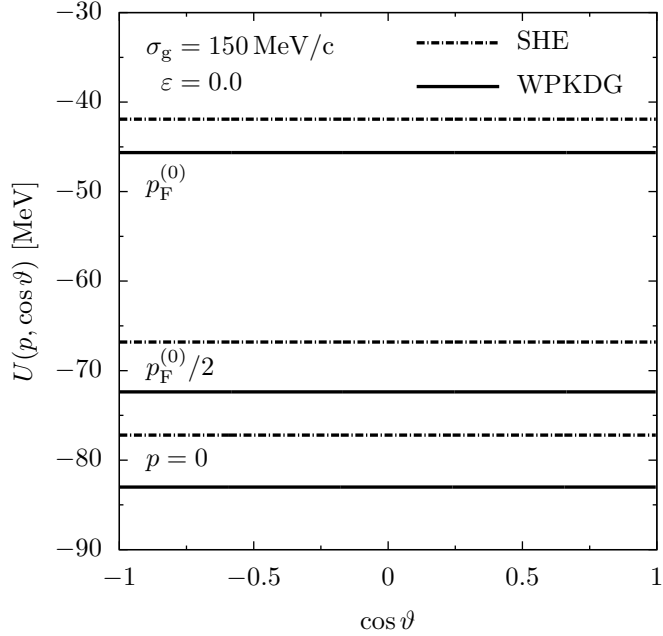}
\caption{\label{fig:cuga00}Optical potential for an isotropic momentum
distribution at~$\rho=2\rho_0$, in the local c.m., for different indicated
momenta~$p$.}
\end{figure}

\begin{figure}
\includegraphics{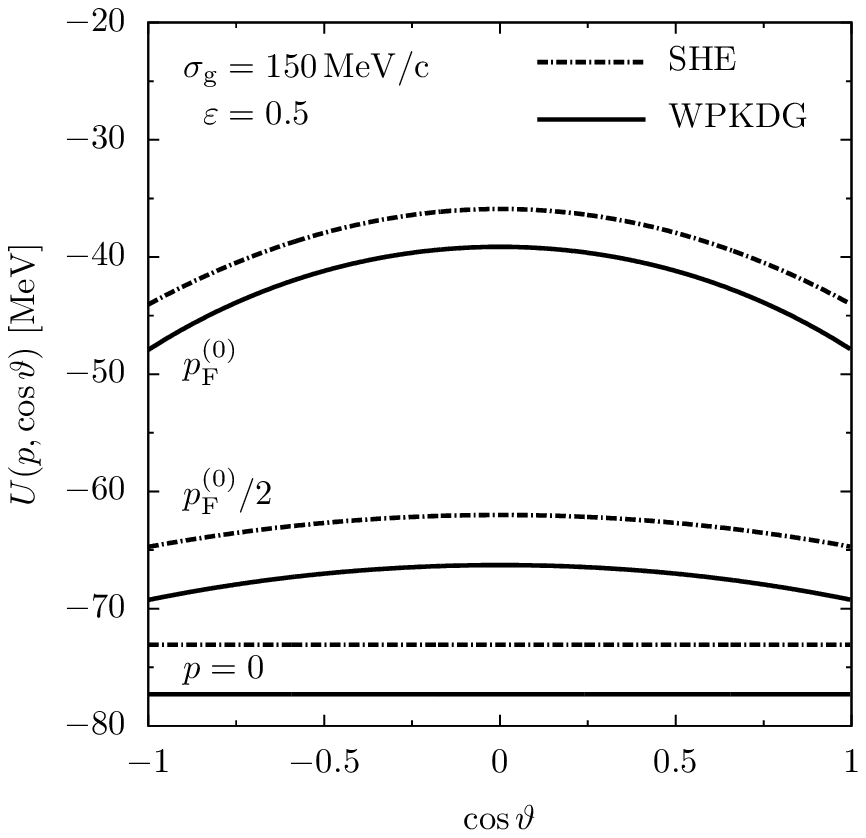}
\caption{\label{fig:cuga05}Optical potential for an anisotropic momentum
distribution with~$\varepsilon=0.5$ at~$\rho=2\rho_0$, in the local c.m., for
different indicated momenta~$p$.}
\end{figure}

\begin{figure}
\includegraphics{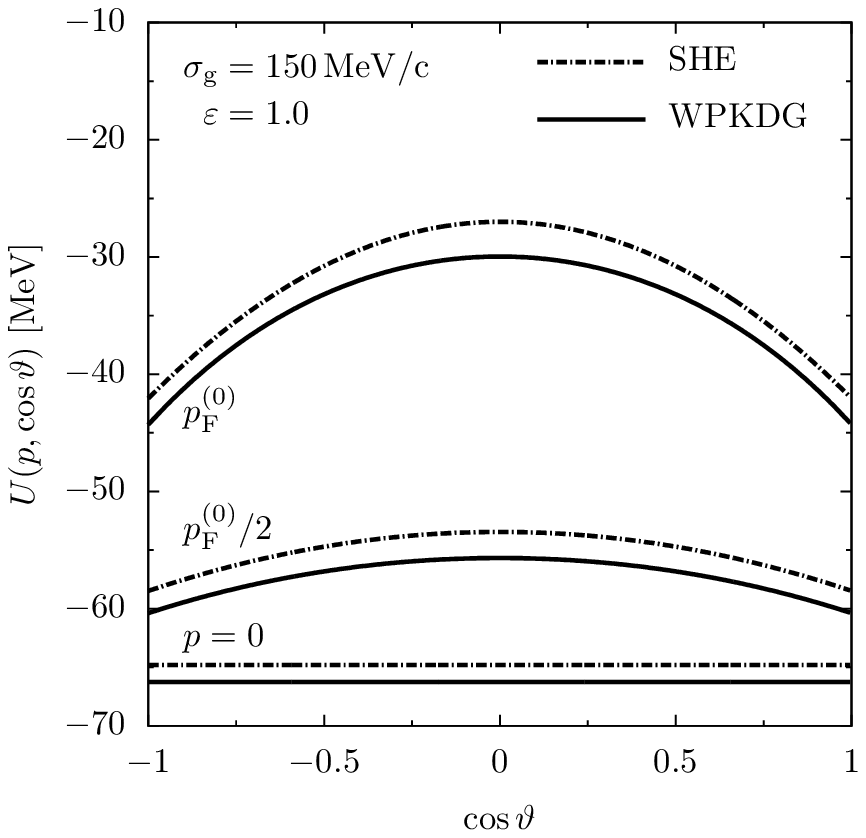}
\caption{\label{fig:cuga10}Optical potential for an anisotropic momentum
distribution with~$\varepsilon=1.0$ at~$\rho=2\rho_0$, in the local c.m., for
different indicated momenta~$p$.}
\end{figure}

\begin{figure}
\includegraphics{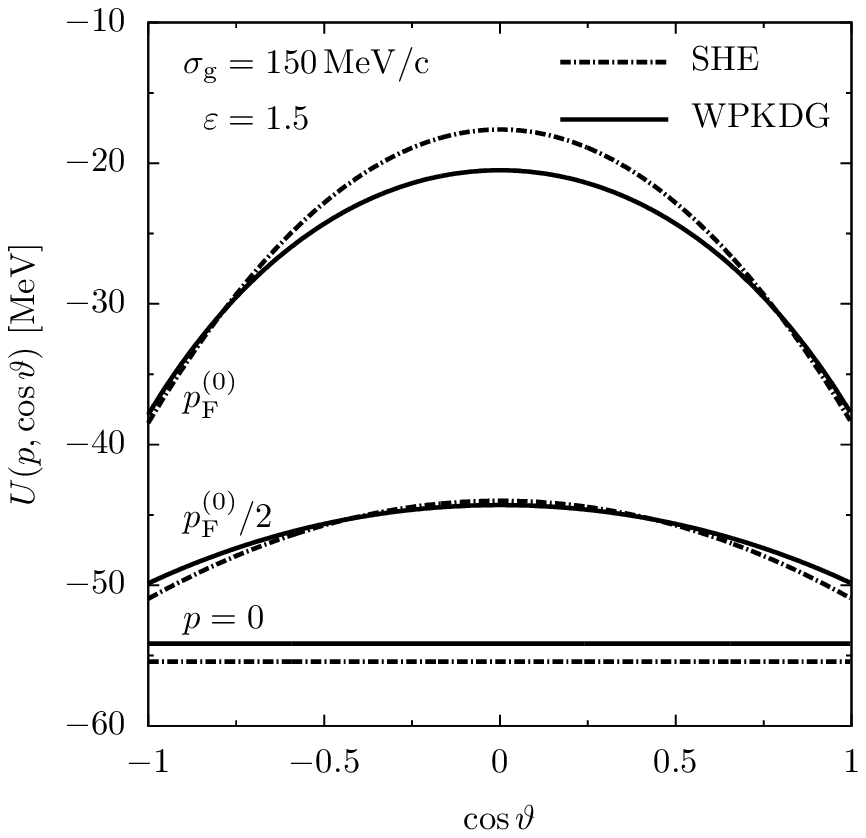}
\caption{\label{fig:cuga15}Optical potential for an anisotropic momentum
distribution with~$\varepsilon=1.5$ at~$\rho=2\rho_0$, in the local c.m., for
different indicated momenta~$p$.}
\end{figure}

To study suitability of SHE for describing situations in excited but not fully
equilibrated matter, we consider results from SHE and WPKDG for a phase-space
density which has the form of an anisotropic Gaussian in momentum:
\begin{widetext}
\begin{equation}
f(\mathbf{r},\mathbf{p})=\frac{\rho(\mathbf{r})}
{(2\pi\sigma_{\text{g}}^2)^{3/2}}\exp\left[-\frac{1}
{2\sigma_{\text{g}}^2}\left(p_{\perp}^2(1+\varepsilon)+\frac{p_{\parallel}^2}
{(1+\varepsilon)^2}\right)\right].\label{eq:gauss}
\end{equation}
\end{widetext}
Here, $\varepsilon$ is a parameter that regulates anisotropy of the momentum
distribution that is axially symmetric about the longitudinal axis. Changes
in~$\varepsilon$ simultaneously change the longitudinal and transverse widths
in such a manner that the density~$\rho$ associated with~$f$ does not change.
Physically allowed values are~$\varepsilon>-1$ and increasing~$\varepsilon$
makes the distribution more prolate.
One needs to relate the parameters~$\varepsilon$ and~$\sigma_{\text{g}}$ to
others, e.g., such as temperature, to determine interesting parameter
ranges. We refer to the kinetic energy density~(\ref{eq:p2m}) to advance in our
considerations. While an isotropic Gaussian~($\varepsilon=0$) yields the
density
\begin{equation}
\left\langle\frac{p^2}{2m}\right\rangle=3\frac{\sigma_{\text{g}}^2}{2m}\rho\;,
\label{eq:p2mgauss}
\end{equation}
for an anisotropic Gaussian~($\varepsilon\neq0$) one obtains
\begin{equation}
\left\langle\frac{p^2}{2m}\right\rangle=\frac{\sigma_{\text{g}}^2}
{2m}\left[\frac{2}{(1+\varepsilon)}+(1+\varepsilon)^2\right]\rho\;.
\label{eq:p2maniso}
\end{equation}
Recalling the Boltzmann model, Gaussian standard deviation~$\sigma$ (for one
direction) and temperature~$T$ are connected via
\begin{equation}
\sigma^2=m_{\text{N}}T\;,\label{eq:sigmat}
\end{equation}
where~$m_{\text{N}}$ is the nucleon mass. Effective temperatures between $15\,
\text{MeV}\leq T\leq 170\,\text{MeV}$ are of interest in heavy-ion collisions.
We assume that, on approach to equilibrium, the distributions are not more
narrow than representing~$T=15\,\text{MeV}$ and not much more spread out than
representing~$T=170\,\text{MeV}$. With this, we arrive at the anisotropy
range~$-0.6\lesssim\varepsilon\lesssim1.5$. In the results that follow
for~$\rho=2\rho_0$, we use the relatively low~$\sigma_{\text{g}}= 150
\,\text{MeV/c}$, which allows for relatively strong momentum dependencies
in~$U$, at low momenta. The subsequent
Figs.~\ref{fig:cuga-05}--\ref{fig:cuga15} show the obtained values of optical
potential both in the SHE and the WPKDG parametrizations, at different
momenta~$p$
(in multiples of ground-state~$p_{\text{F}}^{(0)}$; Sec.~\ref{subsec:rA})
in~$\rho=2\rho_0$ matter with different momentum anisotropies, as a function of
angle relative to the symmetry axis of the distribution.
In the figures, we can see rather good correspondence between the results for
the two parametrizations. While the SHE parametrization explicitly follows the
shape of~$Y_{20}$, it is approximately also the case for WPKDG. Apparently,
higher multipolarity terms in a separable expansion for WPKDG are negligible.
This is likely helped by the smoothness of the momentum distribution. Notably
in heavy-ion collisions the momentum distributions tend to evolve to a smooth
form rather quickly, making the close correspondence between the results likely
in practice.
%%%%%%%%%%%%%%%%%%%%%%%%%%%%%%%%%%%%%%%%%%%%%%%%%%%%%%%%%%%%%%%%%%%%%%%%%%%%%%%
\subsection{\label{subsec:rC}Idealized collision scenario}
\begin{figure}
\includegraphics{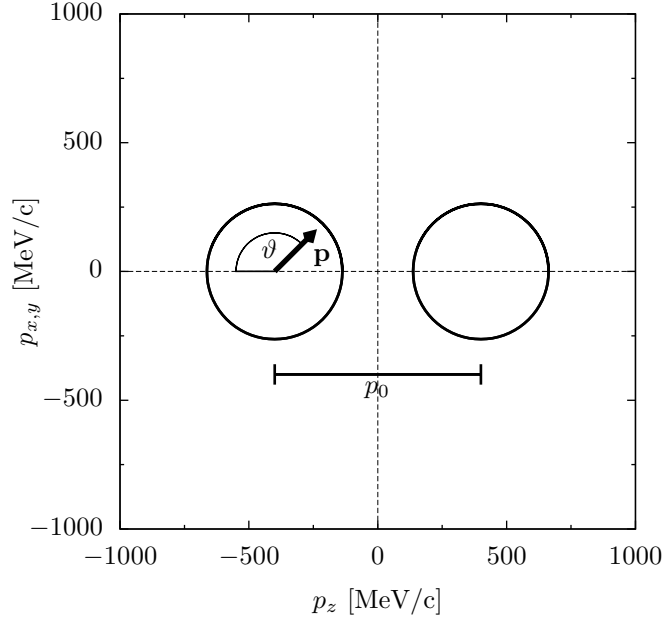}
\caption{\label{fig:cfermileft}Situation within the local rest frame of two
Fermi spheres separated by~$p_0$ in momentum space. The center of the left
sphere serves as point of origin in the first examination of~$U$ for that
situation.}
\end{figure}

\begin{figure}
\includegraphics{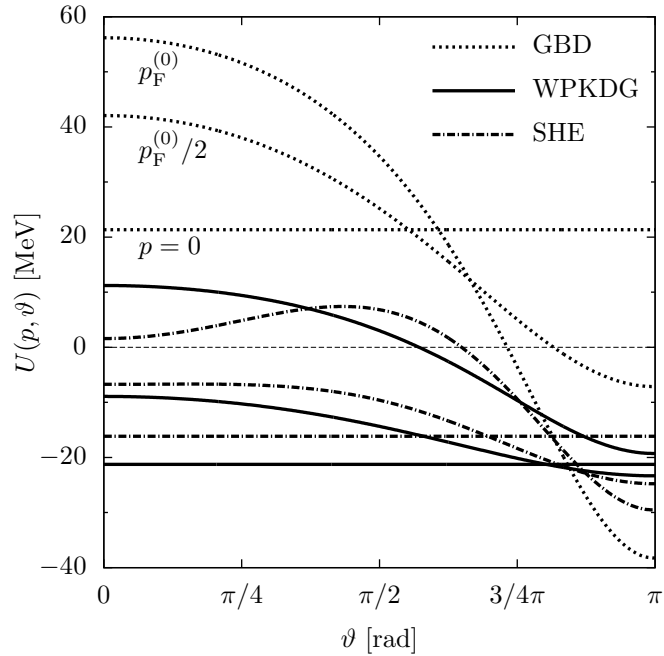}
\caption{\label{fig:cu2ln}Optical potential for the situation represented in
Fig.~\ref{fig:cfermileft}, for different momenta relative to the center of one
of the Fermi spheres, as a function of angle about that center, for the
different parametrizations of the potential.}
\end{figure}

\begin{figure}
\includegraphics{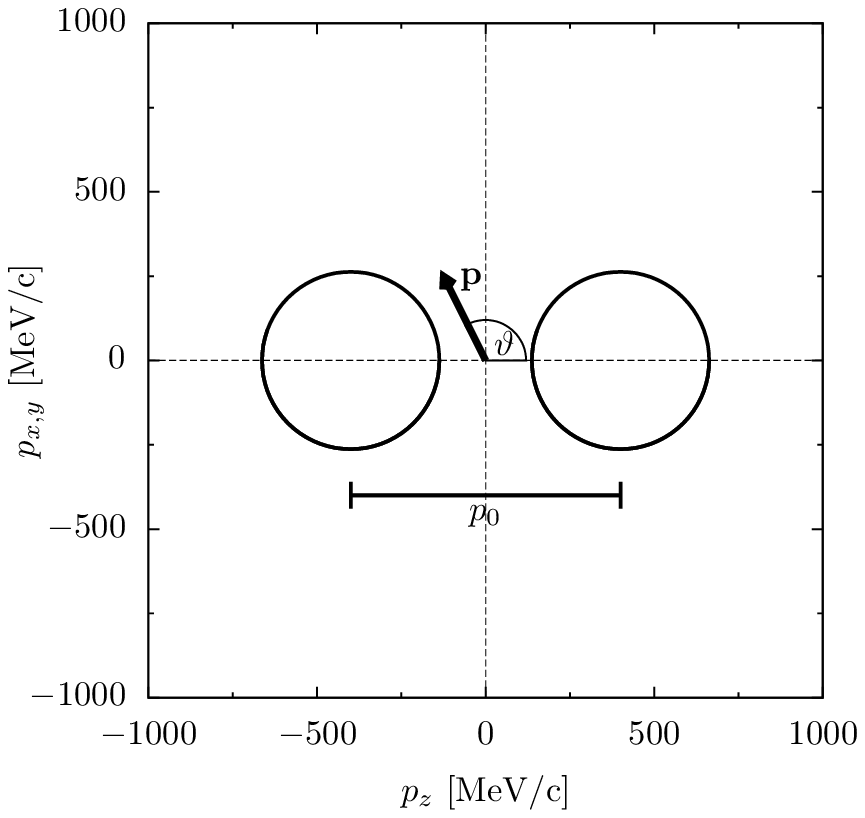}
\caption{\label{fig:cfermicm}Situation within the local rest frame of two Fermi
spheres separated by~$p_0$ in momentum space. The center of mass serves as
point of origin in the second examination of~$U$ for that situation.}
\end{figure}

\begin{figure}
\includegraphics{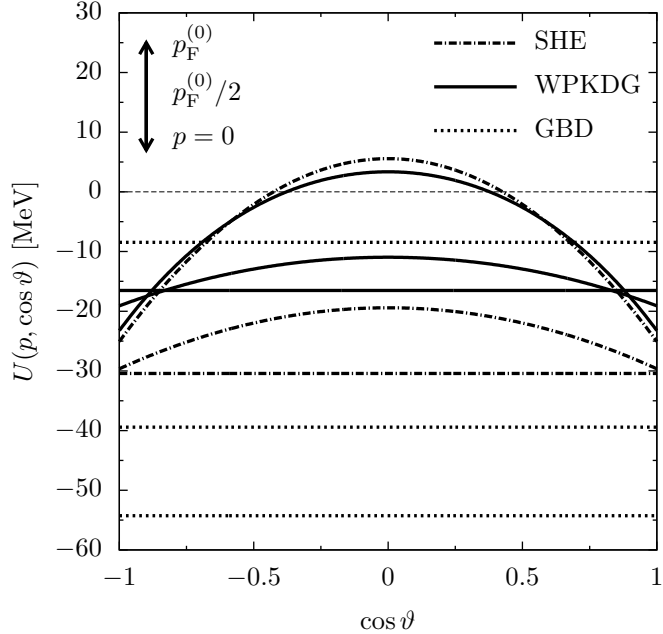}
\caption{\label{fig:cu2cm}Optical potential for the situation represented in
Fig.~\ref{fig:cfermicm}, for different momenta relative to the local center of
mass, as a function of angle about the center, for the different
parametrizations of the potential.}
\end{figure}

The final test situation that we consider, potentially the most challenging for
a description such as SHE, is the early stage of a heavy-ion collision, where
interpenetration of the opposing nuclei has started but no equilibration has
yet occurred. To represent such a situation, Welke \textit{et al.}\ took a
momentum distribution with two Fermi spheres, each representing the
ground state of saturated matter,
\begin{equation}
f(\mathbf{r},\mathbf{p})=\frac{g}{(2\pi\hbar)^3}\,
\theta(p_{\text{F}}^{(0)}-\left|\mathbf{p}\mp\mathbf{p}_0/2\right|)\;,
\label{eq:2fs}
\end{equation}
separated by~$p_0=800\,\text{MeV/c}$ in momentum space. The scenario is
sketched in Fig.~\ref{fig:cfermileft}. To account for relativistic effects in
this colliding nuclear matter scenario, one would rather consider Fermi
ellipsoids than spheres~\cite{Gai99,Fuc03,Gai04}. Such a Lorentz covariant
treatment goes beyond the goals of the present paper.
The optical potentials~$U$ (cf.\ Fig.~\ref{fig:cu2ln}) associated with it are
plotted as functions of the polar angle~$\vartheta$ (defined in
Fig.~\ref{fig:cfermileft}) for different momenta.
The vector momenta~$\mathbf{p}$ are taken with reference to the center of one
sphere, with the angle relative to the symmetry axis of the system. The mean
fields vary not only in overall offset, but also in dependence on angle.
As a point of view additional to that in Welke \textit{et al.}~\cite{Wel88},
presented above, we further consider the mean-field potential at
different~$\mathbf{p}$ relative to the~c.m. as a function of angle~$\vartheta$
about the~c.m.\ (defined in Fig.~\ref{fig:cfermicm}). The SHE, WPKDG, and GBD
potentials
are compared in this fashion in Fig.~\ref{fig:cu2cm}. Significant deviations
can be seen between the potentials at lower momenta~$p$, but lesser at the
highest momentum towards the region where the nucleons are. By construction,
the GBD potential is isotropic in this representation.
The discrepancies at low momenta in Fig.~\ref{fig:cu2cm} show limitations of
the SHE approach. While we chose another phenomenological approach as a
reference
here, rather than a microscopic theory for which sparse nonequilibrium results
exist, we expect a difficulty for SHE, with just a couple of separable terms,
to be universal when momentum distributions change abruptly with momentum.
Fortunately, in heavy-ion collisions the distributions quickly evolve to a
smooth form.
\section{\label{sec:discuss}Discussion and conclusions}
%%%%%%%%%%%%%%%%%%%%%%%%%%%%%%%%%%%%%%%%%%%%%%%%%%%%%%%%%%%%%%%%%%%%%%%%%%%%%%%
\subsection{\label{subsec:buu}Application to transport simulations}
Beyond the scope of this article is the actual application of the anisotropic
mean-field parametrization to reaction simulations, namely BUU calculations.
However, to stress the practical relevance of our work, we
provide a brief prescription of how to employ our results in reactions and name
the benefits one should gain then.

The foundation of BUU methodology is the Boltzmann transport
equation~\cite{Ber84,Dan00} for the single-particle distribution
function~$f(\mathbf{p},\mathbf{r},t)$. To determine the state of the
system one needs to solve the BUU equation for~$f$ in every time step~$t$:
\begin{equation}
\frac{\partial f}{\partial t}+\frac{\partial\epsilon}
{\partial\mathbf{p}}\frac{\partial f}
{\partial\mathbf{r}}-\frac{\partial\epsilon}{\partial\mathbf{r}}\frac{\partial
f}{\partial\mathbf{p}}=I_{\text{coll}}(f)\;. \label{eq:buu}
\end{equation}
The collision integral on the right-hand side of Eq.~(\ref{eq:buu}) accounts for
the time
evolution of the single-particle phase-space density~$f$ evoked by two-body
collisions. The mean field~$U$ enters the Boltzmann equation on the left-hand
side, which takes the motion of particles in the field into account. To find
the single-particle energies~$\epsilon$ one has to take the functional
derivative of the system's net energy~$E$ with respect to the phase-space
density~$f$:
\begin{equation}
\epsilon  = \frac{\delta E}{\delta f} 
= \frac{p^2}{2m}+\frac{\delta V}{\delta f} 
= \frac{p^2}{2m}+U\;.
\end{equation}
If the $p$-dependent part of the mean field is parametrized in terms of an
integral,
\begin{equation}
U(\mathbf{r},\mathbf{p})\propto\int d^{3}p'\,\frac{f(\mathbf{r},\mathbf{p}')}
{1+\left[\frac{\mathbf{p}-\mathbf{p}'}{\Lambda}\right]^2}\;,
\end{equation}
as in the model by Welke \textit{et al.}, one has to determine a different
three-dimensional integral for every momentum~$\mathbf{p}$ and
position~$\mathbf{r}$. When~$f$ is represented in terms of
test particles~\cite{Ber84}, $N_{\text{t}}$~per nucleon, the integration is
replaced with summation over the test particles. With the need to calculate an
integral at the phase-space position of every test particle, the overall effort
in calculating the mean field scales as~$N_{\text{t}}^2$, forcing compromises
at large~$N_{\text{t}}$ countering this rapid growth. However, in the
framework of our SHE model, one would have to evaluate the integrals in the
isotropic term and in~$T^{zz}$ once per time step for a given spatial location.
With this, the effort in calculating the mean field would scale
as~$N_{\text{t}}$, just as for the GBD parametrization or $p$-independent~$U$.
The calculational effort reduced by a factor of the order of~$N_{\text{t}}$,
for the same accuracy, represents a significant advantage of our model over the
WPKDG parametrization.

%%%%%%%%%%%%%%%%%%%%%%%%%%%%%%%%%%%%%%%%%%%%%%%%%%%%%%%%%%%%%%%%%%%%%%%%%%%%%%%
\subsection{\label{subsec:sum}Summary}
We developed a parametrization for the nucleonic mean field~$U$, which
explicitly exhibits an anisotropic behavior for anisotropic phase-space
densities~$f$ and can significantly reduce computational effort and
statistical noise in transport simulations, compared to current practice. On
that account, we made the nuclear energy functional separable in momentum
space, with different terms corresponding to different multipolarity in
spherical angle. As a reference and guideline during the process of setting up
our model, the parametrization by Welke \textit{et al.}~\cite{Wel88} was used.
We evolved step by step the elements of our model by breaking up the
original---and in comparison to our parametrization very costly---convolution
within the potential energy density, substituting it with separable scalar and
tensorial terms, both symmetric in the single-particle properties, and by
taking the functional derivative of~$V$ with respect to~$f$ to obtain~$U$ in a
simple form. In addition to the Skyrme-type parameters in the $\rho$-dependent
part, our model comprises two exclusive parameters, one of
them---$\Lambda_{\text{iso}}$---mainly relevant in isotropic scenarios, the
other one---$\Lambda_{\text{aniso}}$---important for characterizing the
anisotropy of
the mean field. The~latter phenomenological parameters represent range and are
physically expected to be similar. In the framework of our SHE model we can
reasonably well describe cold nuclear matter properties, obtain excellent
results for equilibrated scenarios, and find a good practical agreement for
early stages of a colliding system. The~latter situation has been dealt with in
the literature~\cite{Wel88}. We showed that our ansatz is generally applicable
even when momentum population for the system changes abruptly.

To conclude, we presented a simple way of describing the potential energy
density~$V$ and associated anisotropic momentum-dependent mean field~$U$ in
heavy-ion collisions. Flexibility in coping with interactions for anisotropic
momentum distributions is important because local momentum anisotropies persist
until late in the dynamics of heavy-ion collisions. Exploration of consequences
of the anisotropies, however, should not be stacked against inabilities to
carry out collision calculations, poor statistics, or noise in the calculations.
With the proposed approach, we believe, we resolve these issues.

\begin{acknowledgments}
The authors would like to thank Brent W.~Barker and Jun Hong for discussions.
In particular we would like to express our gratitude to U.~Mosel and
T.~Gaitanos for helpful comments and inspiring suggestions.
This work was supported by National Science Foundation Grant Nos.\ PHY-0800026
and PHY-1068571.
\end{acknowledgments}

\appendix*
\section{\label{app:app}Analytical expressions for ground-state quantities}
When dealing with Fermi spheres in cold-matter scenarios one can find
analytical results for the integrals over a Yukawa interaction kernel,
appearing in Eqs.~(\ref{eq:vwelke}), (\ref{eq:uwelke}), and~(\ref{eq:usep}), for
instance. Thus, one can circumvent many-dimensional numerical integrations.
This obviously does not help in carrying out collision simulations, because the
distributions quickly depart from the Fermi spheres. Particularly useful are
the following identities:
\begin{widetext}
\begin{eqnarray}
\label{eq:vanalytic}
\int_0^{p_{\text{F}}}\int_0^{p_{\text{F}}} d^{3}p\> d^{3}p'\frac{1}
{1+\left[\frac{\mathbf{p}-\mathbf{p}'}{\Lambda}\right]^2}=\frac{32\pi^2}
{3}p_{\text{F}}^4\Lambda^2\Bigg[&~&\frac{3}{8}-\frac{\Lambda}
{2p_{\text{F}}}\arctan\frac{2p_{\text{F}}}{\Lambda}-\frac{\Lambda^2}
{16p_{\text{F}}^2}\nonumber\\
&+&\left(\frac{3}{16}\frac{\Lambda^2}{p_{\text{F}}^2}+\frac{1}
{64}\frac{\Lambda^4}{p_{\text{F}}^4}\right)\ln\left(1+\frac{4p_{\text{F}}^2}
{\Lambda^2}\right)\Bigg],
\end{eqnarray}
\begin{eqnarray}
\label{eq:wanalytic}
\int_0^{p_{\text{F}}} d^{3}p'\frac{1}{1+\left[\frac{\mathbf{p}-\mathbf{p}'}
{\Lambda}\right]^2}=\pi\Lambda^3\Bigg[&~&\frac{p_{\text{F}}^2+\Lambda^2-p^2}
{2p\Lambda}\ln\frac{(p+p_{\text{F}})^2+\Lambda^2}{(p-
p_{\text{F}})^2+\Lambda^2}+\frac{2p_{\text{F}}}{\Lambda}\nonumber\\
&-&2\left(\arctan\frac{p+p_{\text{F}}}{\Lambda}-\arctan\frac{p-p_{\text{F}}}
{\Lambda}\right)\Bigg].
\end{eqnarray}
\end{widetext}
\newpage
\bibliography{paper}

\end{document}